\documentclass[a4paper,11pt]{article}
\usepackage{pos}
\usepackage{wrapfig}
\newcommand{\ct}{c_t}
\newcommand{\ctt}{\tilde{c}_t}

\newcommand{\SigmaNPH}{\sigma_{\rm NP}^{H}}
\newcommand{\Sigmakt}{\bar\sigma_{\kt}}
\newcommand{\Sigmaktt}{\bar\sigma_{\ktt}}
\newcommand{\kt}{\kappa_t}
\newcommand{\ktt}{\tilde{\kappa}_t}
\newcommand{\cts}{\ct}
\newcommand{\ctp}{\ctt}

\newcommand{\Sigmact}{\bar\sigma_{\ct}}
\newcommand{\Sigmactt}{\bar\sigma_{\ctt}}
\newcommand{\Sigmactctt}{\bar\sigma_{\ct,\ctt}}
\title{Virtual corrections  to top-pair production from new physics}
\usepackage{caption}
\author*[a]{Simone Tentori}

\affiliation[a]{Center for Cosmology, Particle Physics and Phenomenology, Université Catholique de Louvain, Louvain-la-Neuve, Belgium}

\emailAdd{simone.tentori@uclouvain.be}

\abstract{New physics and SM parameters can be studied and constrained by looking at the modifications to top-pair differential kinematical distributions due to off-shell effects. I present here three case studies: the determination of the Higgs couplings to the top-quark, a search for generic BSM scalar and  pseudoscalar states,  and a search for Axion Like Particles (ALP). 
The corresponding models have been implemented in the \texttt{UFO} format to allow automatic computations  within MadGraph5\_aMC@NLO.}
\FullConference{42nd International Conference on High Energy Physics (ICHEP2024)\\
18-24 July 2024\\
Prague, Czech Republic\\}

\begin{document}
\maketitle

\section{Introduction}
Top-pair production is one of the most studied processes at the LHC. On the theoretical side differential kinematical distribution can be computed at next-to-next-to-leading order (NNLO) precision in Quantum Chromodynamics (QCD) \cite{Czakon:2015owf, Catani:2019hip} and at  next-to-leading order (NLO) precision in electroweak (EW)  \cite{Pagani:2016caq,Czakon:2017wor}. 
On the experimental side, Run III of LHC in combination with HL-LHC will produce more than 3 billions $t\bar t$-pairs, and this large statistics together with an expected reduction of systematics will lead to an unprecedented precision in the measurement of the top-pair kinematical distributions. 

\begin{figure}[ht]
    \centering
    \includegraphics[width=0.55\linewidth]{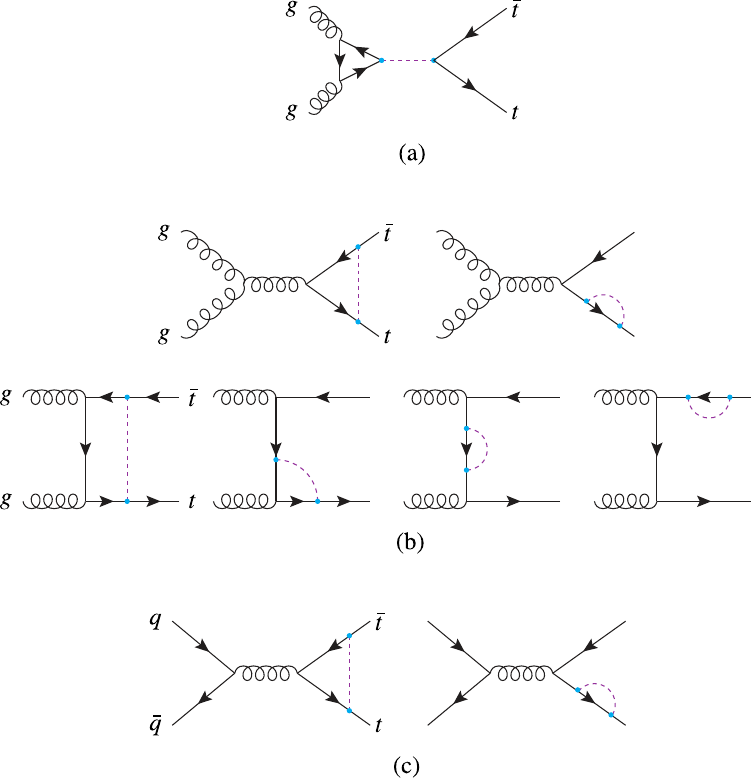}
    \caption{One-loop Feynman diagrams for $t\bar t$ production obtained by the exchange of a virtual scalar.}
    \label{fig:diagrams}
\end{figure}

Given the current and expected precision even very subtle effects, such as those coming from the exchange of virtual particles, can be studied. They  affect the kinematical distributions not only inclusively but also at the differential level. Here we will consider the case originated by the virtual exchange of spin-zero particles, and in particular, those described by the Feynman diagrams shown in Fig. \ref{fig:diagrams} interfering with the usual tree-level SM $t\bar t$-pair production diagrams.
In all cases considered, the one-loop contributions present a Sommerfeld enhancement close to the $t\bar t$ threshold which can help  to solve tension in that region via off-shell New Physics (NP) effects.

The scalar state entering the Feynman diagrams can be either a SM particle (such as the Higgs boson) which couplings to the top-quark we want to measure or a completely new Beyond Standard Model (BSM) state. In the following sections I will analyse  different possibilities in the three scenarios. All models presented here have been implemented in the \texttt{UFO}\cite{Degrande:2011ua} format to allow computations in MadGraph5\_aMC@NLO\cite{Alwall:2014hca} and can be requested to the author.

\section{Top yukawa determination}
\begin{figure}[ht]
    \centering
    \includegraphics[width=0.38\linewidth]{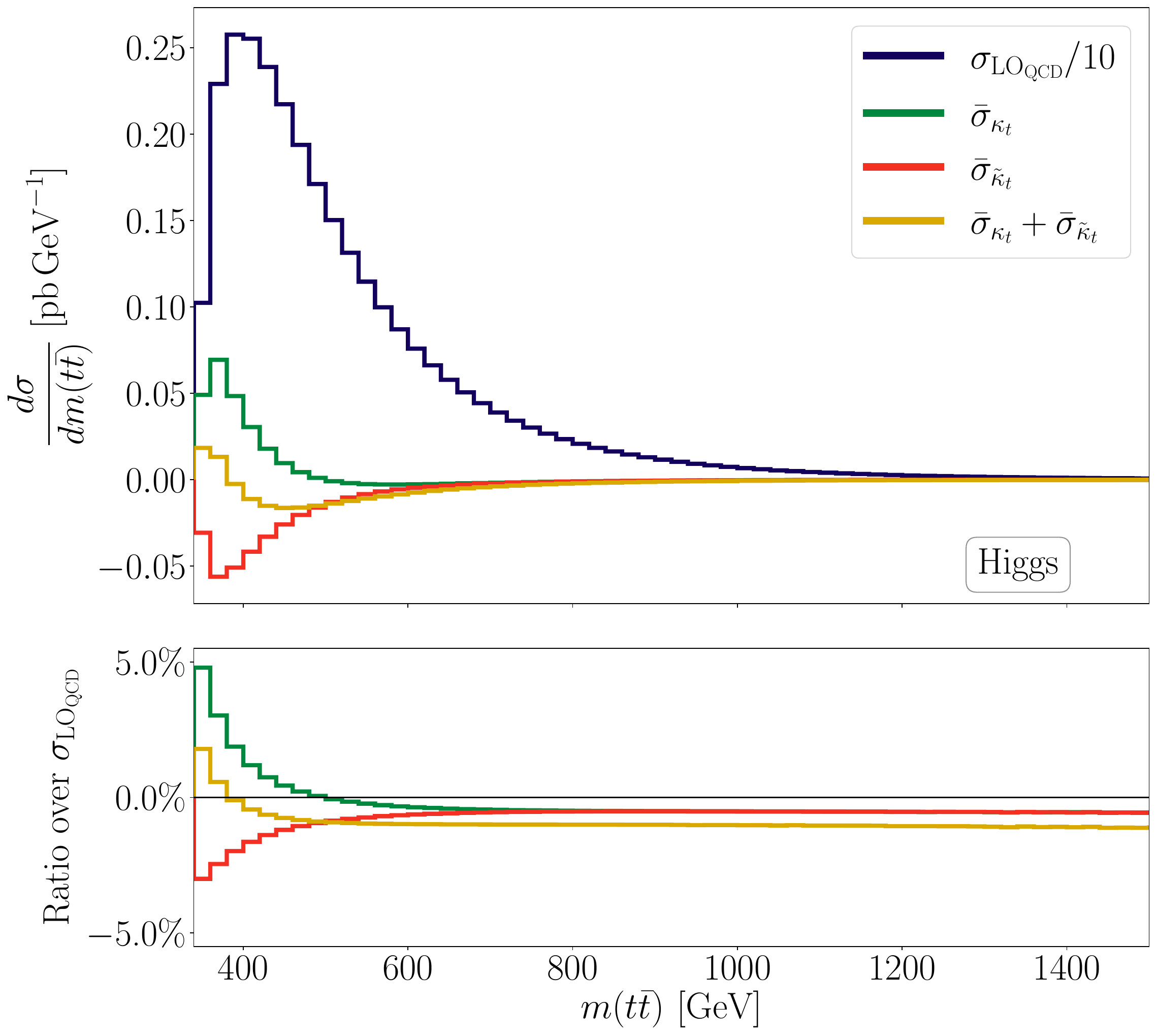}
    \includegraphics[width=0.38\linewidth]{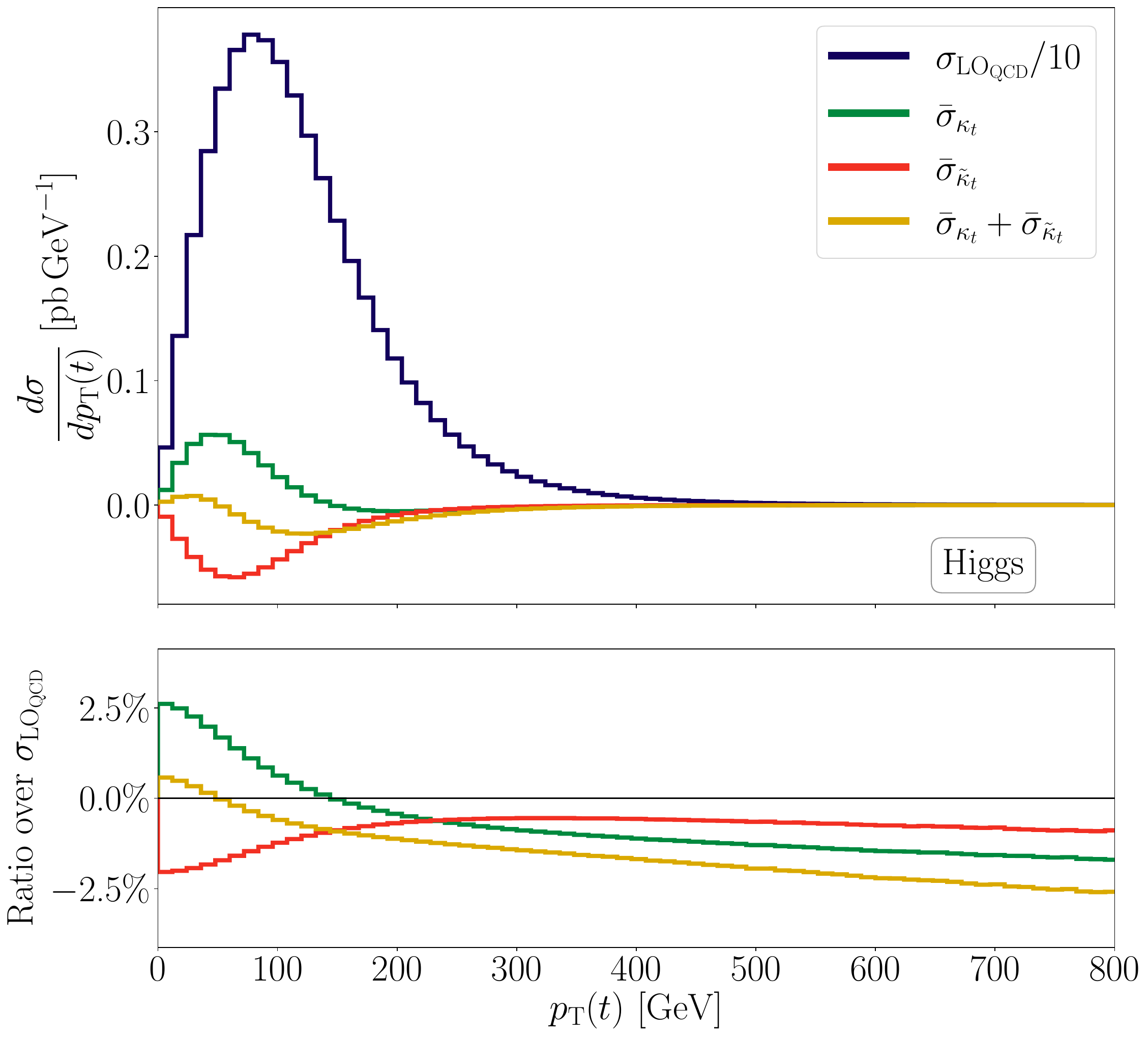}
    \caption{In the main panels: $\sigma_{{\rm LO}_{\rm QCD}}$ divided by ten (blue) and the loop corrections $\SigmaNPH$ evaluated at different $(\kt, \ktt)$ benchmarks: $(1,0)$ green, $(0,1)$  red and $(1,1)$ mustard. In the insets the ratio over $\sigma_{{\rm LO}_{\rm QCD}}$ is plotted.}
    \label{fig:Higgsdistr}
\end{figure}

Higgs couplings modifications w.r.t. SM can be captured in the $\kappa$ framework. In its general form the Higgs boson interaction with the top quark can be rewritten as
\begin{align}
    \mathcal{L}_{Ht\bar t}&= -\frac{y_t}{\sqrt{2}}\bar t (\kappa_t+i\tilde\kappa_t \gamma_5)t H=\mathcal{L}_{\rm SM}+\mathcal{L}_{\rm NP},\label{eq:kappa}
\end{align}
where
\begin{align}
 \mathcal{L}_{\rm SM}&=-\frac{y_t}{\sqrt{2}}\bar t t H,\qquad \mathcal{L}_{\rm NP}=-\frac{y_t}{\sqrt{2}}\bar t [(\kappa_t-1)+i\tilde\kappa_t \gamma_5]t H\,,
\end{align}
i.e., we have separated the SM part of the Lagrangian from the one containing the NP effects.
The NP virtual corrections to the $t\bar t$-differential distribution arising from the diagrams in Fig. \ref{fig:diagrams} can be parameterized as 
\begin{equation}
\SigmaNPH=(\kt^2-1)\, \Sigmakt + \ktt^2 \, \Sigmaktt\,, \label{eq:sigmanpinkts}\,
\end{equation}
and are plotted for the top-pair invariant mass and top-quark transverse momentum distributions in Fig. \ref{fig:Higgsdistr}, we point out that CP-breaking terms $\kt\ktt$ are not present since vanishing.

We observe that at the threshold the pure CP-even (green) and the pure CP-odd (red) configuration give opposite sign contributions, leading to a partial cancellation in the CP-mixed (mustard) contribution.

The two $\kt$ and $\ktt$ parameters can be fitted separately or together, this method has been used for the top-yukawa determination from the CMS collaboration \cite{CMS:2019art,CMS:2020djy} using the implementation in \cite{Martini:2021uey}, which however, as far as we know, does not include the s-channel diagram in Fig \ref{fig:diagrams}. The contributions obtained considering the full virtual corrections and the ones obtained excluding the $s$-channel diagrams are almost identical if we limit ourselves to the CP-even interaction, but they are much different if a CP-odd or a CP-mixed interaction is taken into consideration (and thus will give very different parameters uncertainties in a simultaneous $\kt-\ktt$ fit \cite{Maltoni:2024wyh}).

\begin{figure}
\begin{minipage}[b]{0.4\linewidth}
\renewcommand{\arraystretch}{1.5}
    \centering
    \begin{tabular}{l|c|c}
       &  ${\kappa_t}^{+1\sigma,\,2\sigma,\,3\sigma}_{-1\sigma,\,2\sigma,\,3\sigma}$ & ${\tilde\kappa{_t}}^{+1\sigma,\,2\sigma,\,3\sigma}_{-1\sigma,\,2\sigma,\,3\sigma}$ \\[1.5pt]
              \hline
               LHC & $1.00^{+0.28,\,0.52,\,0.72}_{-0.41,\,1.0,\,1.0}$&$0.0^{+0.59,\,1.05,\,1.43}_{-0.59,\,1.06,\,1.44}$\\

    \end{tabular}
    \captionof{table}{Best fit and corresponding errors values for $\kappa_t$ and $\tilde\kappa_t$ obtained in a separate fit of the two modifiers. }
    \label{tab:Higgscalarfit}
     \end{minipage}
     \hfill
     \begin{minipage}[b]{0.52\linewidth}
\renewcommand{\arraystretch}{1.5}
    \centering
\includegraphics[width=0.95\textwidth]{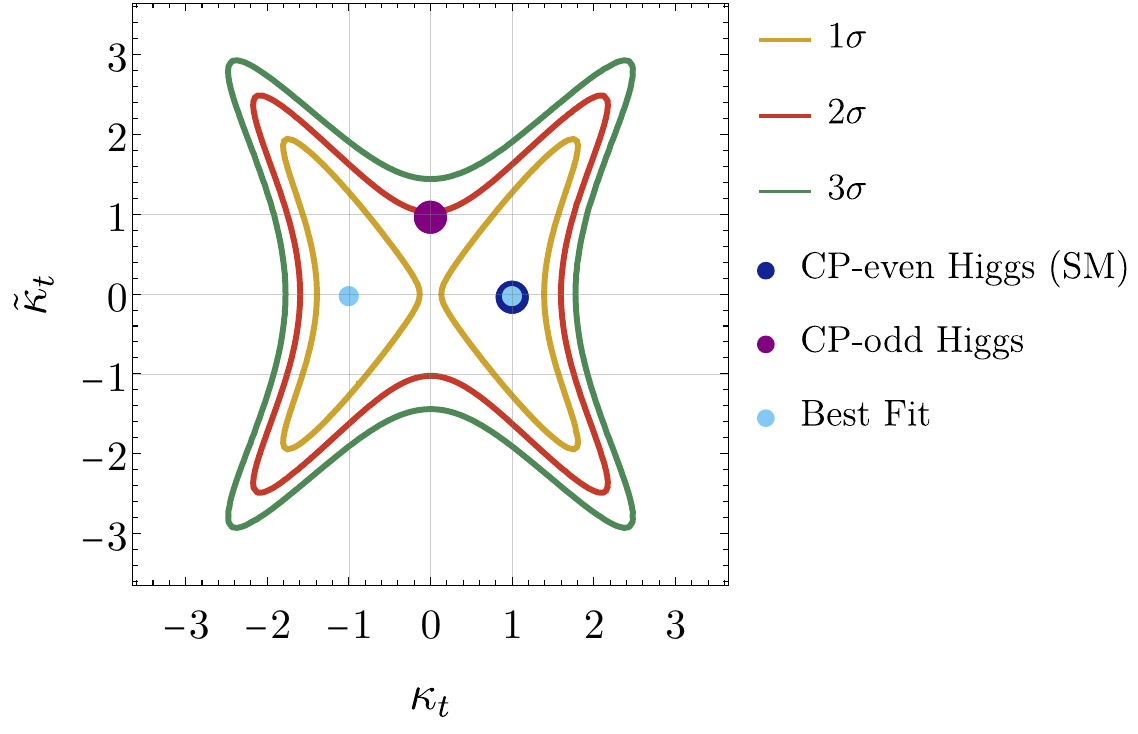}
    \captionof{figure}{Contours at the $1\sigma, 2\sigma, 3\sigma$ confidence levels for the simultaneous fit of the $\kt-\ktt$ modifiers.}
    \label{fig:Higgscalarfit}
     \end{minipage}
\end{figure}

To test what precision in the Higgs coupling determination can be achieved with such a method we have performed a fit separately in  $\kt$ and $\ktt$, details on the experimental data, uncertainties and theoretical calculations can be found in \cite{Maltoni:2024wyh}. The results of these fits are reported in Tab. \ref{tab:Higgscalarfit}. This method is particularly well suited for determining the CP nature of the Higgs boson. When fitting the CP-odd modifier we exactly obtain $\ktt=0$. This is related to opposite threshold signs of the corrections stemming from the CP-even and CP-odd interaction. We also performed a simultaneous fit of the $\kt-\ktt$ modifiers, the results are shown in Fig. \ref{fig:Higgscalarfit}, the CP-odd configuration of the Higgs boson is almost excluded at $2 \sigma$, CP-mixed configurations (in particular along the $\kt-\ktt$ diagonal) are instead more difficult to exclude due to the partial cancellations discussed above. 

\section{Arbitrary CP top-philic scalar}
\begin{figure}[htb]
\begin{minipage}[b]{0.55\linewidth}
    \centering
    \includegraphics[width=0.5\linewidth]{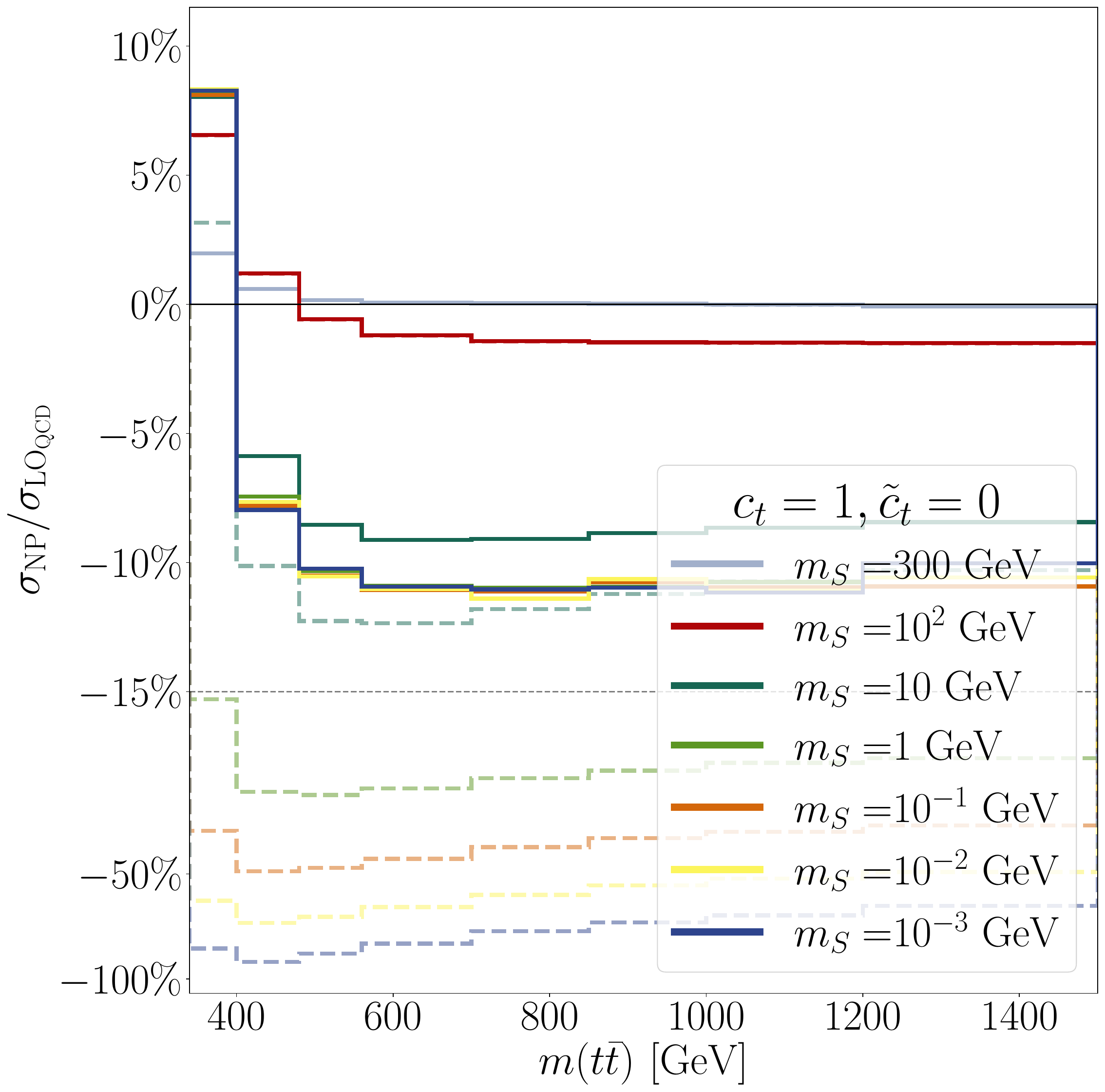}
    \includegraphics[width=0.49\linewidth]{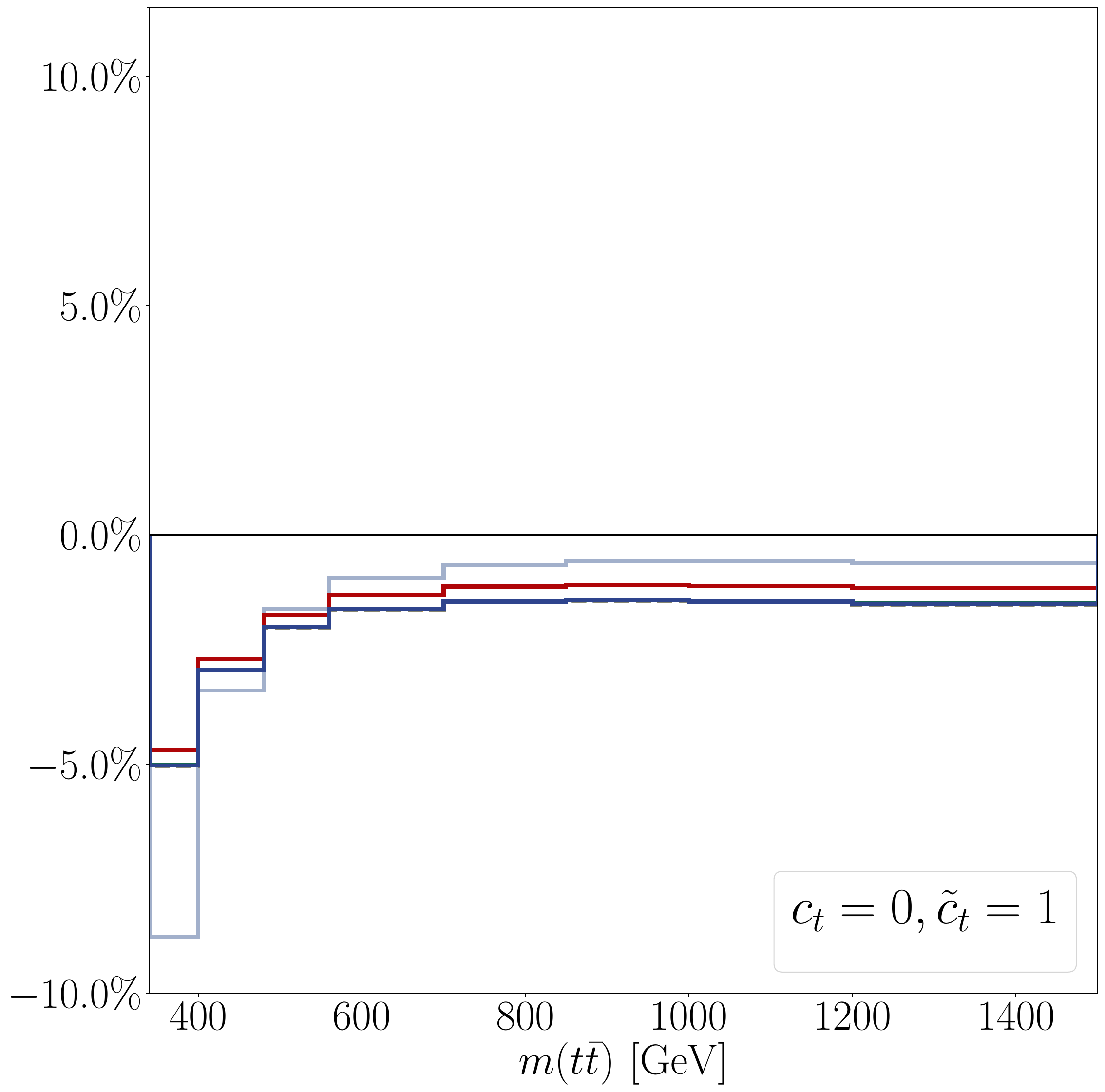}

    \captionof{figure}{From left to right: the purely scalar, the purely pseudoscalar . The solid lines are obtained by adding the contribution from $S$ real emission with a $p_{\rm T}(S)<20$ GeV cut. }
    \label{fig:ScalarMasses}
\end{minipage}
 \begin{minipage}[b]{0.45\textwidth}
     \centering
    \includegraphics[width=\linewidth]{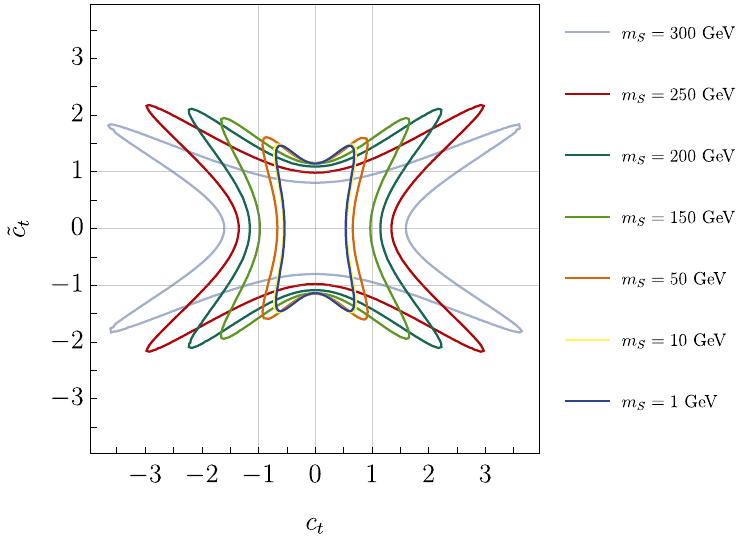}
     \captionof{figure}{Bound at $2\sigma$ in the $\ct-\ctt$ plane }
     \label{fig:chimass}
 \end{minipage}
\end{figure}

A case very similar to the Higgs presented before is the search of a top-philic CP arbitrary-scalar $S$ whose Lagrangian can be written as
\begin{equation}
  \mathcal{L}_{S}\equiv \frac{1}{2}\partial_\mu S\partial^\mu S-\frac{1}{2}m_S^2 S^2 -\bar t (\cts+i\ctp\gamma_5)\psi_t S \label{eq:intS}\,.   
\end{equation}
The correction to the $t\bar t$-pair differential distributions can be parameterized as 
\begin{equation}
 \sigma_{\rm NP}\equiv\Sigmact \ct^2+\Sigmactt \ctt^2+\Sigmactctt \ct \ctt  \,.
\end{equation}
Also in this case we will find that  $\Sigmactctt=0$. The virtual corrections to top-pair production for different values of $m_S$ are plotted in Fig. \ref{fig:ScalarMasses}.
 For all masses, the corrections obtained in the CP-even case ($\ctt=0$) at the threshold have opposite sign with respect to the CP-odd case ($\ct=0$). In Fig. \ref{fig:ScalarMasses} the dashed lines correspond to the pure virtual corrections, while the solid lines include the contribution from the final state $t\bar t S$ with a $p_{\rm T}(S)<20$ GeV cut. 
 The inclusion of the real emission corrects every benchmark in which $\ct\neq 0$. This is because virtual corrections coming from a scalar exchange $S$ coupling with a CP-even interaction contain soft singularities; a quite surprising result is instead the absence of these singularities in the CP-odd case. Adding the real emission does not change at all the results obtained with the pure virtual corrections. We have computed the expected sensitivity in the $\ct-\ctt$ plane for a possible LHC search using this approach. The results are shown in Fig. \ref{fig:chimass}. When the mass $m_S$ changes we see very different shapes in the contour, in particular for higher masses the pseudoscalar coupling $\ctt$ is more excluded, while the exclusion on $\ct$ becomes stronger for lower masses.

\section{Top-philic Axion Like Particles}

Axion Like Particles (ALP) at colliders have recently raised particular attention in the collider community. A particular compelling scenario is the so-called top-philic ALP, \footnote{More literature references can be found in the cited papers.} the Lagrangian coupling the ALP $a$ to the top-quark reads

\begin{equation}
   \mathcal{L}_{\rm int}=-\frac{c_t}{2f_a}(\partial_\mu a)\bar t \gamma^\mu \gamma_5 t\, .
\end{equation}

\begin{figure}[ht]
    \centering
    \includegraphics[width=0.625\linewidth]{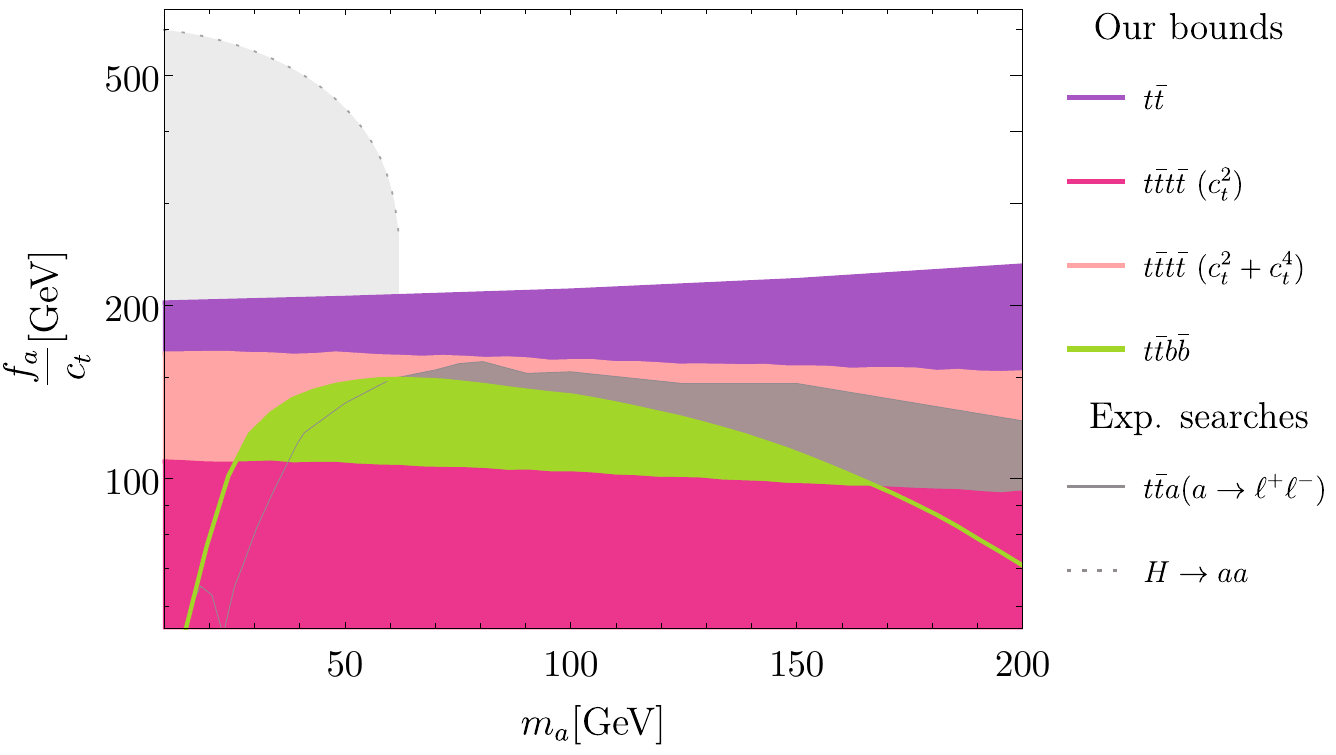}
    \caption{Exclusion plot for the top-philic ALP coupling, on the y-axis the quantity $f_a/c_t$. The indirect constraint obtained with the method described in this paper is the strongest.}
    \label{fig:ALP}
\end{figure}

This interaction is only similar to the one of a pseudoscalar particle, but there are some major differences, in particular, the axion decay into di-photons is super-suppressed, making this particle extremely difficult to look for at the LHC\cite{Blasi:2023hvb}. Given the failure of direct searches, the contribution obtained from ALP virtual exchange in $t\bar t$-pair production can be used to try to constraint the parameter space \cite{Blasi:2023hvb, Phan:2023dqw}. A comprehensive exclusion plot for the top-philic ALP coupling $c_t/f_a$ is shown in Fig. \ref{fig:ALP}, given the elusiveness of this particle the indirect constraint obtained via off-shell virtual correction is actually the strongest constraint obtainable.

\section{Conclusions}
I have illustrated the main characteristics and opportunities of virtual correction to top-pair production showing how they can constraints both SM and NP couplings, giving the best bound for particularly elusive NP cases. All the models here presented have been implemented in the \texttt{UFO} format for computations in MadGraph5\_aMC@NLO.
\bibliographystyle{JHEP}
\bibliography{bibref}

\end{document}